\documentclass{article}

\usepackage{amsthm}
\usepackage{url}
\usepackage{authblk}
\usepackage{amsmath,amstext,amssymb,amsfonts,verbatim}

\usepackage{color}

\usepackage{graphicx,algorithmic,algorithm,rotating}
\allowdisplaybreaks[1]

\newenvironment{myenumerate}{
\begin{enumerate}
 \setlength{\itemsep}{1pt}
 \setlength{\parskip}{0pt}
 \setlength{\parsep}{0pt}}{\end{enumerate}
}

\begin{document}

\title{\vspace{-4cm}Algorithmic Information Dynamics of Persistent Patterns and Colliding Particles\\in the Game of Life\thanks{Source code available at: \protect\url{https://github.com/hzenilc/algorithmicdynamicGoL.git}. An online implementation of estimations of graph complexity is available at \protect\url{http://www.complexitycalculator.com}}}

\author{Hector Zenil$^{1,2,3,4}$, Narsis A. Kiani$^{1,2,3,4}$, Jesper Tegn\'er$^{2,3,5}$\\
$^1$ Algorithmic Dynamics Lab, Centre for Molecular Medicine,\\ Karolinska Institute, Stockholm, Sweden\\
$^2$ Unit of Computational Medicine, Department of Medicine,\\Karolinska Institute, Stockholm, Sweden \\
$^3$ Science for Life Laboratory, SciLifeLab, Stockholm, Sweden\\
$^4$ Algorithmic Nature	Group, LABORES	for	the	Natural	and\\Digital Sciences, Paris, France\\
$^5$ Biological and Environmental Sciences and Engineering Division,\\	Computer, Electrical and Mathematical	Sciences and Engineering\\Division,	King Abdullah University of Science and \\Technology (KAUST), Kingdom	of Saudi Arabia\\
{\{hector.zenil, narsis.kiani, jesper.tegner\}@ki.se}}

\date{}
\maketitle{}

\begin{abstract}Without loss of generalisation to other systems, including possibly non-deterministic ones, we demonstrate the application of methods drawn from \textit{algorithmic information dynamics} to the characterisation and classification of emergent and persistent patterns, motifs and colliding particles in Conway's Game of Life (GoL), a cellular automaton serving as a case study illustrating the way in which such ideas can be applied to a typical discrete dynamical system. We explore the issue of local observations of closed systems whose orbits may appear open because of inaccessibility to the global rules governing the overall system. We also investigate aspects of symmetry related to complexity in the distribution of patterns that occur with high frequency in GoL (which we thus call \textit{motifs}) and analyse the distribution of these motifs with a view to tracking the changes in their algorithmic probability over time. We demonstrate how the tools introduced are an alternative to other computable measures that are unable to capture changes in emergent structures in evolving complex systems that are often too small or too subtle to be properly characterised by methods such as lossless compression and Shannon entropy. \\

\noindent{}\textbf{Keywords:} Kolmogorov-Chaitin complexity; cellular automata; algorithmic probability; algorithmic Coding Theorem, Turing machines; Algorithmic Information Theory; Game of Life; dynamic pattern classification
\end{abstract}

\section{Introduction}

It has been proven that there are quantitative connections between indicators of algorithmic information content (or algorithmic complexity) and the chaotic behaviour of dynamical systems that is related to their sensitivity to initial conditions.
Some of these results and the relevant references are, for example, given in~\cite{vieri}. Previous numerical approaches, such as the one used in~\cite{vieri} and others cited in the same paper, including those proposed by the authors of the landmark textbook on Kolmogorov complexity~\cite{li}, make use of computable measures, in particular measures based on popular lossless compression algorithms, and suggest that non-computable approximations cannot be used
in computer simulations or in the analysis of experiments.
One of the aims of this paper is to prove that a new measure~\cite{methodszenil,maininfopaper,zenilgraph} based on the concept of algorithmic  probability, that has been shown to be more powerful~\cite{emergence,bdm} than computable approximations~\cite{zenildata} such as popular lossless compression algorithms (e.g. LZW), can overcome some previous limitations and difficulties in profiling orbit complexity, difficulties
particularly encountered in the investigation of the behaviour of local observations typical of computer experiments in, e.g., cellular automata research. This is because, for example, typically-used popular lossless compression algorithms are closer to Shannon entropy in their operation~\cite{emergence} than to a measure of algorithmic complexity, and Shannon entropy is not only limited in that it can only quantify statistical regularities, but it is also not robust and can easily be fooled in very simple ways~\cite{zenilphysres}.

The concept of \textit{Algorithmic Information Dynamics} (or simply \textit{algorithmic dynamics}) was introduced in~\cite{maininfopaper} and draws heavily on the theories of Computability and Algorithmic Information. It is a calculus with which to study the change in the causal content of a dynamical system's orbits when the complex system is perturbed or unfolds over time. We demonstrate the application and utility of these methods in characterising evolving emergent patterns and interactions (collisions) in a well-studied example of a dynamical (discrete) complex system that has been proven to be very expressive by virtue of being computationally universal~\cite{durand}.

The purpose of \textit{algorithmic dynamics} is to trace in detail the changes in algorithmic probability---estimated by local observations—-produced by natural or induced perturbations in evolving open complex systems. This is possible even for partial observations that may look different but that come from the same source. For in general, we can only have partial access in the real-world to a system's underlying generating mechanism, yet from partial observations algorithmic models can be derived, and their likelihood of being the producers of the phenomena observed estimated.

\subsection{Emergent patterns in the Game of Life}

\begin{figure}[htbp!]
  \centering
    A\hspace{2.3cm}B\hspace{2.3cm}C\\
\includegraphics[width=7.8cm]{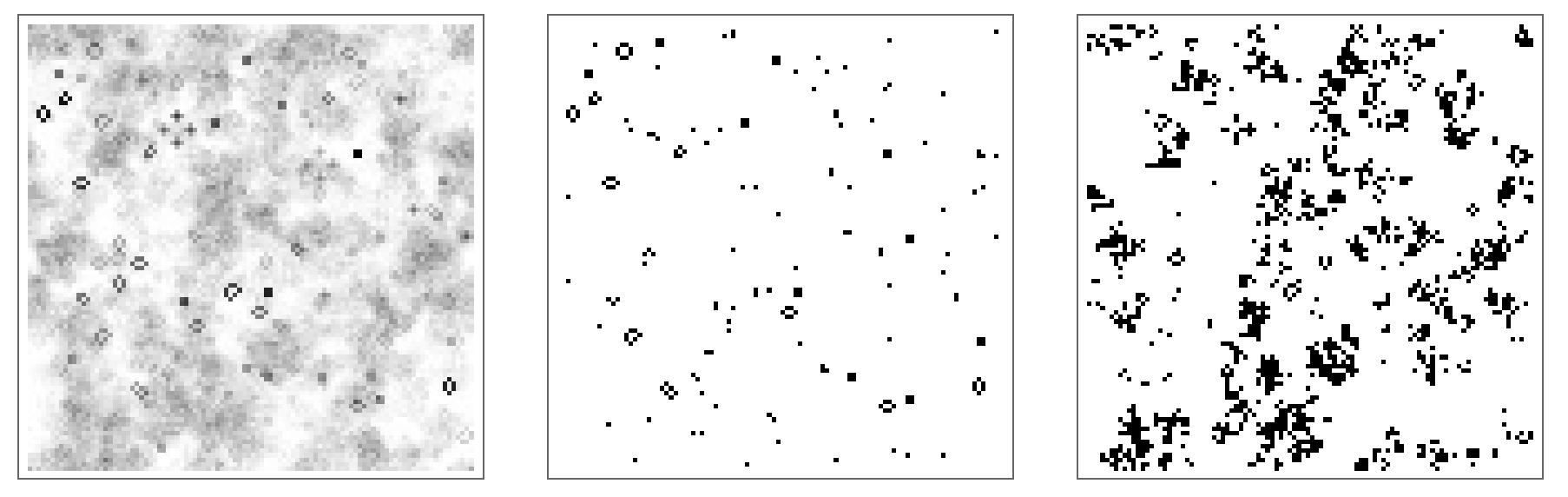}
  \caption{\label{gol}A typical run of the Game of Life (GoL). A: Density plot with persistent motifs highlighted and vanishing ones in various lighter shades of grey. B: Only prevalent motifs from the initial condition as depicted in C.} 
\end{figure}

Conway's Game of Life~\cite{conway} (GoL) is a 2-dimensional cellular automaton (see Figure~\ref{gol} Sup. Inf.). A cellular automaton is a computer program that applies in parallel a global rule composed of local rules on a tape of cells with symbols (e.g. binary). The local rules governing GoL are traditionally written as follows:

\begin{myenumerate}
\item A live cell with fewer than two live neighbours dies.
\item A live cell with more than three live neighbours dies.
\item A live cell with two or three live neighbours continues to live.
\item A dead cell with three live neighbours becomes a live cell.
\end{myenumerate}

Each of these is a local rule governing a special case, while the set of rules 1-4 constitute the global rule defining the Game of Life.

Following~\cite{durand}, we call a configuration in GoL that contains only a finite number of `alive' cells and prevails a \textit {pattern}. If such a pattern occurs with high frequency we call it a \textit{motif}.

For example, so-called \textit{`gliders'} are a (small) pattern that emerges in GoL with high frequency. The most frequent glider motif (see Fig.~\ref{GoLcomps3}D) travels diagonally at a speed of $t/4$ across the grid and is the smallest and fastest motif in GoL, where $t$ is the automaton runtime from initial condition $t=0$.

Glider collisions and interactions can produce other particles such as so-called `blocks', `beehives', `blinkers', `traffic lights', and a less common pattern known as the `eater'. Particle collisions in cellular automata, as in high particle physics supercolliders, have been studied before~\cite{martinez}, demonstrating the computational capabilities of such interactions where both annihilation and new particle production is key. Particle collision and interaction profiling may thus be key in controlling the way in which computation can happen within the cellular automaton. For example, using only gliders, one can build a pattern that acts like a finite state machine connected to two counters. This has the same computational power as a universal Turing machine, so using the glider, the Game of Life automaton was proven to be Turing-universal, that is, as powerful as any computer with unlimited memory and no time constraints~\cite{berlekamp}.

GoL is an example of a 2-dimensional cellular automaton that is not only Turing-universal but also intrinsically universal~\cite{durand}. This means that the Game of Life not only computes any computable function but can also emulate the behaviour of any other 2-dimensional cellular automaton (under rescaling).

\section{Preliminaries and Methods}

We are interested in applying some measures related to (algorithmic) information theory to track the local dynamical changes of patterns and motifs in GoL that may shed light on the local but also the global behaviour of a discrete dynamical system, of which GoL is a well-known case study. To this end, we compare and apply Shannon Entropy; Compress, an algorithm implementing lossless compression; and a measure related to and motivated by algorithmic probability (CTM/BDM) that has been used in other contexts with interesting results.

\subsection{Shannon entropy}

The entropy $Η$ of a discrete random variable $s$ with possible values ${s_1, \dots, s_n}$ and probability distribution $P(s)$ is defined as:

    $$H(s)=-\sum_{i=1}^n P(s_i) \log_2 P(s_i)$$

\noindent where if $P(s_i) = 0$ for some $i$, then $log_2(0)=0$.

In the case of arrays or matrices $s$ is a random variable in a set of arrays or matrices according to some probability distribution (usually the uniform distribution is assumed, given that Shannon entropy per se does not provide any means or methods for updating $P(s)$).

\subsection{Lossless compression}

Lossless compression algorithms have traditionally been used to approximate the Kolmogorov complexity of an object. Data compression can be viewed as a function that maps data onto other data using the same units or alphabet (if the translation is into different units or a larger or smaller alphabet, then the process is called a 're-encoding' or simply a 'translation'). Compression is successful if the resulting data are shorter than the original data plus the decompression instructions needed to fully reconstruct said original data. For a compression algorithm to be lossless, there must be a reverse mapping from compressed data to the original data. That is to say, the compression method must encapsulate a bijection between ``plain" and ``compressed" data, because the original data and the compressed data should be in the same units.

A caveat about lossless compression: lossless compression based on the most popular algorithms such as LZW (Gzip, PNG, Compress) that are traditionally considered to be approximations to algorithmic (Kolmogorov) complexity are closer to Shannon entropy than to algorithmic complexity (which we will denote by $K$). This is because these popular lossless compression algorithms implement a method that traverses the object of interest looking for statistical repetitions from which a basic grammar is produced based entirely on their frequency of appearance. This means that common lossless compression algorithms overlook many algorithmic aspects of data that are invisible to them because they do not produce any statistical mark.

\subsection{Algorithmic probability and complexity}

Algorithmic Probability is a seminal concept in the theory of algorithmic information. The algorithmic probability of a string $s$ is a measure that describes the probability that a valid (not part of the beginning of any other) random program $p$ produces the string $s$ when run on a universal Turing machine $U$. In equation form this can be rendered as 

$$m(s) = \sum_{p:U(p) = s} 1/2^{|p|}$$

That is, the sum over all the programs $p$ for which $U$ outputs $s$ and halts.

The Algorithmic Probability~\cite{solomonoff,levin} measure $m(s)$ is related to algorithmic complexity $K(s)$ in that $m(s)$ is at least the maximum term in the summation of programs, given that the shortest program carries the greatest weight in the sum. The Coding Theorem further establishes the connection between $m(s)$ and $K(s)$ as follows: 

\begin{equation}
\label{codingtheorem}
|-\log_2 m(s) - K(s)| < c
\end{equation}

\noindent where $c$ is a fixed constant independent of $s$. The Coding Theorem implies that~\cite{d4,d5} one can estimate the algorithmic complexity of a string from its frequency by rewriting Eq.~\ref{codingtheorem} as: 

\begin{equation}
\label{ctm}
K_m(s)=-\log_2 m(s) + c
\end{equation}

\noindent where $\mathcal{O}(1)$ is a constant. One can see that it is possible to approximate $K$ by approximations to $m$ (such finite approximations have also been explored in~\cite{finite} on integer sequences), with the added advantage that $m(s)$ is more sensitive to small objects~\cite{d4} than the traditional approach to $K$ using lossless compression algorithms, which typically perform poorly for small objects (e.g. small patterns).

A major improvement in approximating the algorithmic complexity of strings, images, graphs and networks based on the concept of algorithmic probability (AP) offers different and more stable and robust approximations to algorithmic complexity by way of the so-called algorithmic Coding theorem (c.f. below). The method, called the Coding Theorem Method, suffers some of the same drawbacks as other approximations to $K$, including lossless compression, related to the additive constant involved in the \textit{invariance theorem} as introduced by Kolmogorov, Chaitin and Solomonoff~\cite{kolmo,chaitin,solomonoff} that guarantees convergence towards $K$ at the limit without the rate of convergence ever being known. The chief advantage of the algorithm is, however, that algorithmic probability (AP)~\cite{solomonoff,levin} not only looks for repetitions but for algorithmic causal segments, such as in the deterministic nature of the digits of $\pi$, without the need for wild assumptions about the underlying mass distributions.

\begin{figure}[ht]
  \centering
  \includegraphics[width=11.3cm]{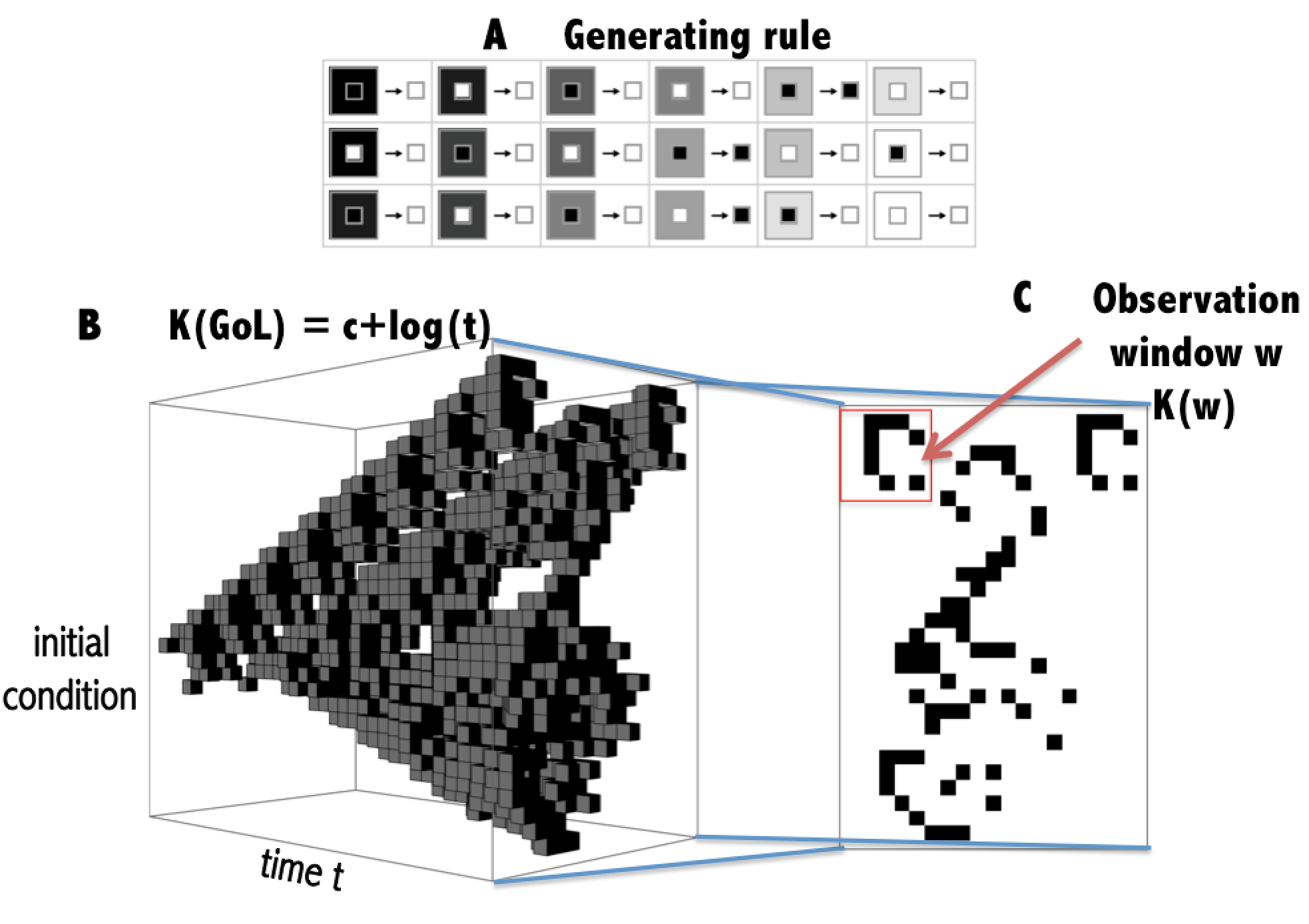}
  
 \caption{\label{obs}The algorithmic complexity of an observation. A: Generating rule of Conway's Game of Life (GoL), a 2-dimensional Cellular Automaton whose global rule is composed of local rules that can be represented by the average of the values of the cells in the (Moore) neighbourhood (a property also referred to as 'totalistic'~\cite{nks}). B: 3D space-time representation of successive configurations of GoL after 30 steps. C: Projected slice window $w$ of an observation of the evolution of B, the last step of GoL.} 
\end{figure}

As illustrated in Figure~\ref{obs}, an isolated observation window does not contain all the algorithmic information of an evolving system. In particular, it may not contain the complexity to be able to infer the set of local generating rules, and  hence the global rule of a deterministic system (Figure~\ref{obs}A). So in practice the phenomena in the window appear to be driven by external processes that are random for all practical purposes, while some others can be explained by interacting/evolving local patterns in space and time (Figure~\ref{obs}C). This means that even though GoL is a fully deterministic system and thus its algorithmic complexity $K$ can only grow by $log(t)$ (Figure~\ref{obs}B), one can meaningfully estimate $K(w)$ of a cross-section $w$ (Figure~\ref{obs}C) of an orbit of a deterministic system like GoL and study its algorithmic dynamics (the change of $K(w)$ over time).

\subsection{Coding Theorem and Block Decomposition Methods}

The method studied and applied here was first defined in~\cite{kolmo2d,bdm}, and is in many respects independent of the observer to the greatest possible extent. For example, unlike popular implementations of lossless compression used to approximate algorithmic complexity (such as LZW), the method based on Algorithmic Probability averages over a large number of computer programs that were found to accurately (without loss of any information) reproduce the output, thus making the problem of the choice of enumeration less relevant, as against the more arbitrary choice of a particular lossless compression algorithm, especially one that is mostly a variation of limited measures such as Shannon entropy. The advantage of the measure of graph algorithmic complexity is that when it diverges from algorithmic complexity—-because it requires greater computational power—-it can only behave as poorly as Shannon entropy~\cite{bdm}, but any behaviour divergent from Shannon entropy can only be an improvement on entropy and a more accurate estimation of the actual information contained in the object based on local calculations of algorithmic complexity.

The \textit{Coding Theorem Method} (CTM)~\cite{d4,d5} is rooted in the relation established by Algorithmic Probability between frequency of production of a string from a random program and its Kolmogorov complexity (Eq.~\ref{codingtheorem}, also called the algorithmic \textit{Coding theorem}, in contrast with the Coding theorem in classical information theory). Essentially, it uses the fact that the more frequent a string (or object), the lower its algorithmic complexity; and strings of lower frequency have higher algorithmic complexity. As has been said, BDM actually calculates Shannon Entropy combined with better approximations, by way of local estimations, of algorithmic complexity.

The approach adopted here consists in determining the algorithmic complexity of a matrix by quantifying the likelihood that a random Turing machine operating on a 2-dimensional tape can generate it and halt. The \textit{Block Decomposition Method} (BDM) then decomposes the matrix into smaller matrices for which we can numerically calculate the algorithmic probability by running a large set of small 2-dimensional deterministic Turing machines, and upon application of the algorithmic Coding theorem, its algorithmic complexity. Then the overall complexity of the original matrix is the sum of the complexity of its parts, albeit with a logarithmic penalisation for repetitions, given that $n$ repetitions of the same object only adds $\log_2 n$ complexity to its overall complexity, as one can simply describe a repetition in terms of the multiplicity of the first occurrence. More formally, the Kolmogorov complexity of a matrix $G$ is defined as follows:

\begin{equation}
\label{newecaeq}
BDM(g,d) = \sum_{(r_u,n_u)\in A(G)_{d\times d}} \log_2(n_u)+CTM(r_u)
\end{equation}
where $K_m(r_u)$ is the approximation of the algorithmic (Kolmogorov-Chaitin) complexity of the subarrays $r_u$ arrived at by using the algorithmic Coding theorem (Eq.~\ref{ctm}), a method that we denote by CTM, and $A(G)_{d\times d}$ represents the set with elements $(r_u,n_u)$, obtained when decomposing the matrix of $G$ into non-overlapping squares of size $d$ by $d$. In each $(r_u,n_u)$ pair, $r_u$ is one such square and $n_u$ its multiplicity (number of occurrences). From now on $K_{BDM} (g,d=4)$ will be denoted only by $K(G)$, but it should be taken as an approximation to $K(G)$ unless otherwise stated (e.g. when speaking of the theoretical true $K(G)$ value).

The only parameters used for the decomposition of BDM as suggested in~\cite{bdm} were the maximum 12 for strings and 4 for arrays, given the current best CTM approximation~\cite{d5} based on an empirical distribution based on all Turing machines with up to 5 states, and no string/array overlapping decomposition for maximum efficiency (as it runs in linear time) and for which the error (due to boundary conditions) is bounded~\cite{bdm}.

An advantage of these algorithm-based measures is that the 2-dimensional versions of both CTM and BDM are native bidimensional measures of complexity and thus do not destroy the 2-dimensional structure of a matrix. This is achieved by making a generalisation of the algorithmic Coding theorem using 2-dimensional Turing machines. In this way we can define the probability of production of a matrix as the result of a randomly chosen deterministic 2-dimensional-tape Turing machine without any array transformations of a string making it dependent on an arbitrary mapping.

\section{Experiments and Numerical Results}

\subsection{Algorithmic probability of emergent patterns}

\begin{figure}[htbp!]
  \centering
  A\hspace{5cm}B\\
  
  \medskip
  
\includegraphics[width=4.8cm]{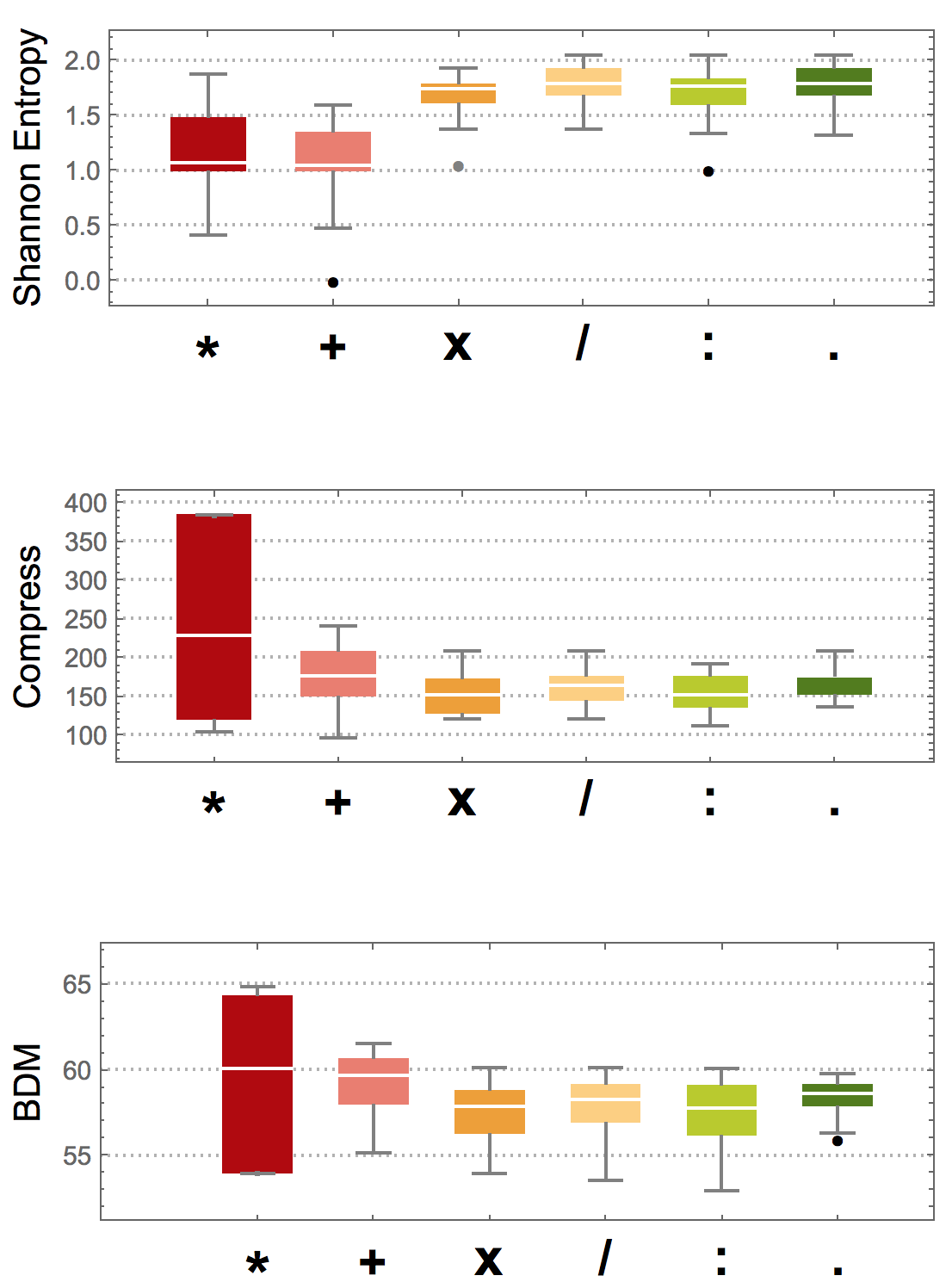}\hspace{.5cm}
    \includegraphics[width=5cm]{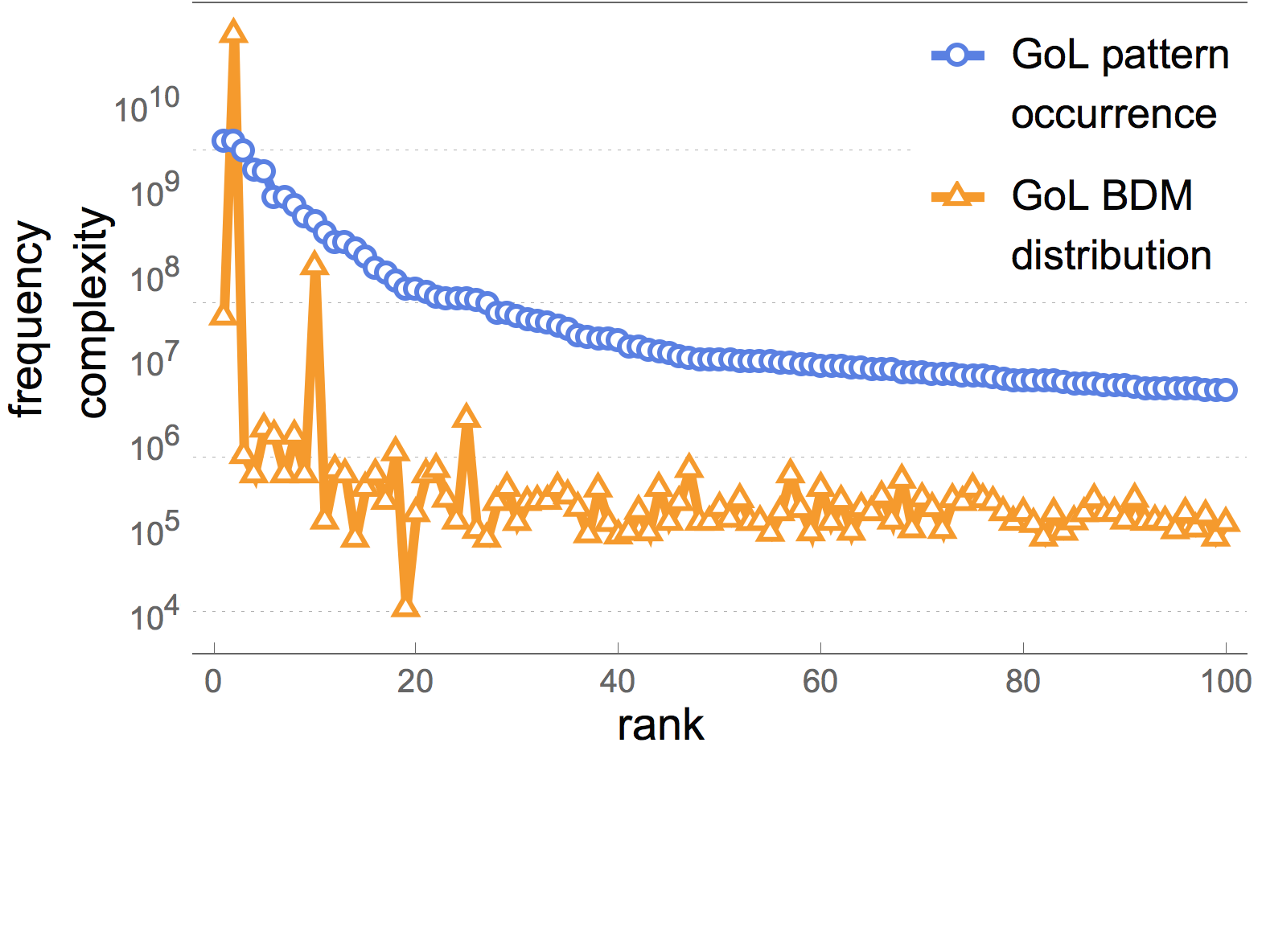}\\
  
  \medskip
  
  C\hspace{5.5cm}D\\
   
   \medskip
   
  \includegraphics[width=11.3cm]{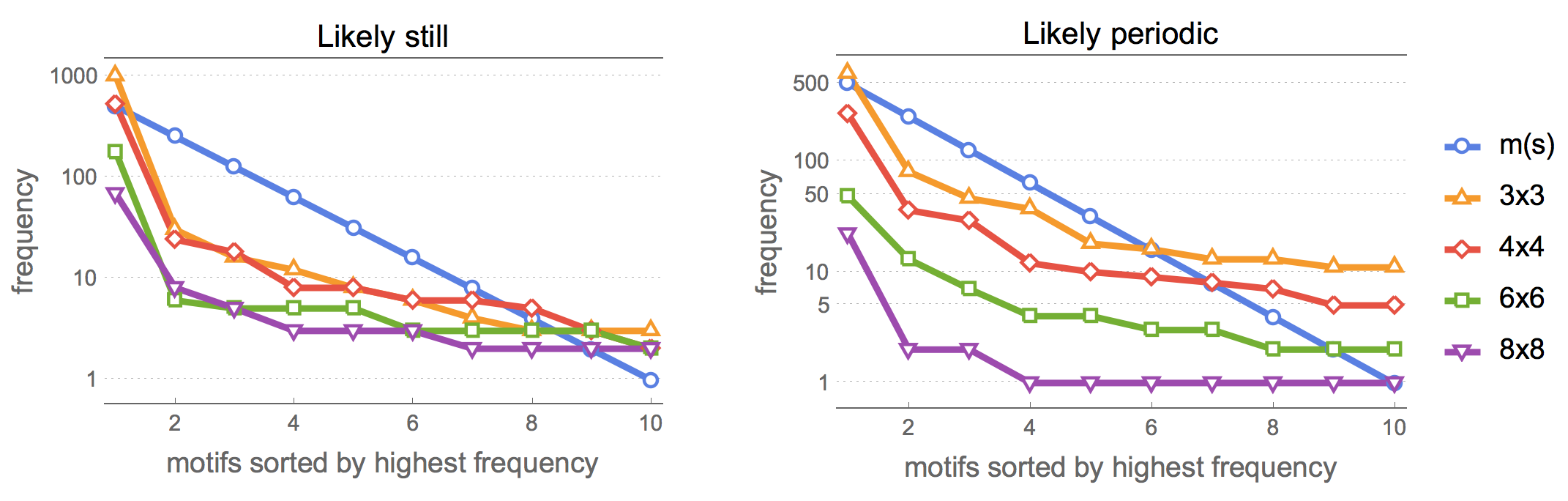}
  
  \caption{\label{dist}A: Classical and algorithmic measures versus symmetries of the top 100 most frequent patterns (hence motifs) in GoL. The measures show diverse (and similar) abilities to separate patterns with the highest and lowest number of symmetries. Notation for the square dihedral group $D_4$: invariant to all possible rotations (*), to all reflections (+), to 2 rotations (X) only and to 2 reflections (/), 1 rotation (:) and 1 reflection (.). B: The heavily long-tail distribution of local persistent patterns in GoL (of less than $10x10$ pixels) from the 100 most frequent emerging patterns and of (C and D) most-likely still and periodic structures.} 
\end{figure}

Figure~\ref{dist}A suggests that highly symmetric patterns/motifs that produce about the same number of black and white pixels and look similar (small standard variation) for Entropy can actually have more complex shapes than those collapsed by Entropy alone. Similar results were obtained before and after normalising by pattern size (length $\times$ width). Symmetries considered include the square dihedral group $D_4$, i.e. those invariant to rotations and reflections. Shannon entropy characterises the highest symmetry as having the lowest randomness, but both lossless compression and algorithmic probability (BDM) suggest that highly symmetric shapes can also reach higher complexity. 

The distribution of motifs (the 100 most frequent local persistent patterns, also called \textit{ash} as they are the debris of random interactions) of GoL are reported in \url{http://wwwhomes.uni-bielefeld.de/achim/freq_top_life.html} by starting from 1\,829\,196 random seed (in a torus configuration) with initial density 0.375 black cells over a grid size of $2048 \times 2048$ and from which 50\,158\,095\,316 objects were found. 

Given the structured nature of the output of GoL, taking larger blocks reveals this structure (see Figure~\ref{dist}B-D). If the patterns were statistically random the block decomposition would display high block entropy values, and the distributions of patterns would look more uniform for larger blocks. However, larger blocks remain highly non-uniform, indicating a heavy tail, as is consistent with a distribution corresponding to the algorithmic complexity of the patterns--that is, the simpler the more frequent. Indeed, the complexity of the patterns can explain 43\% (according to a Spearman rank correlation test, $p$-value $8.38\times10^{-6}$) of the simplicity bias in the distribution of these motifs (see Fig.~\ref{dist}B). 

\begin{figure}[htbp!]
  \centering
  A\hspace{5cm}\\

\medskip
 
  \includegraphics[width=9.5cm]{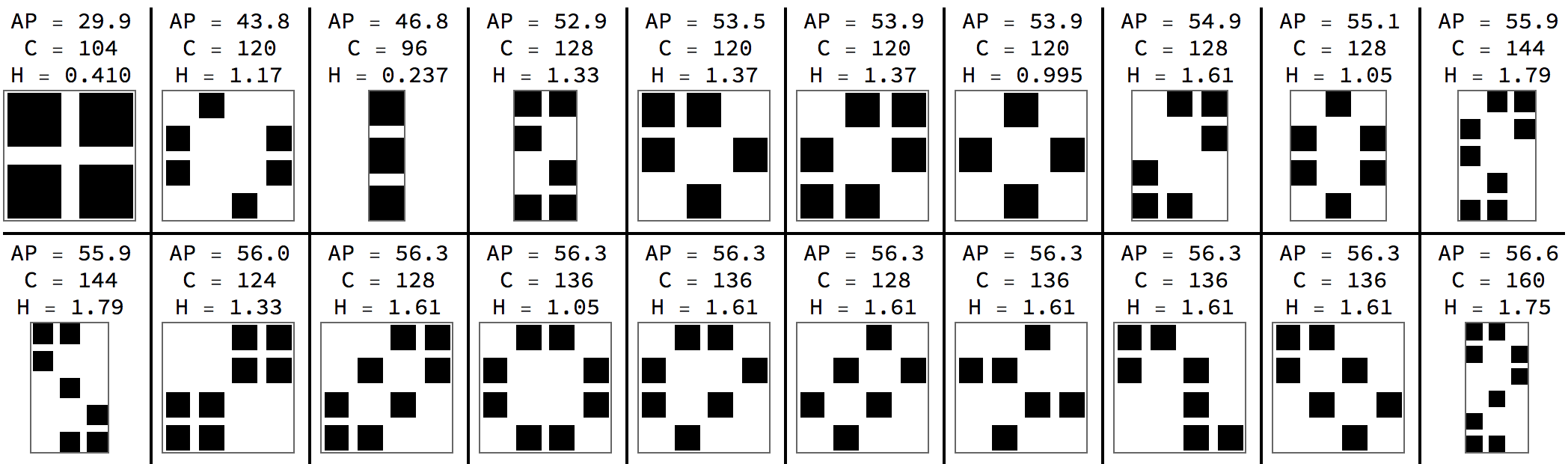}\\
 
 \medskip

  B\hspace{5cm}\\
  
  \medskip

  \includegraphics[width=10cm]{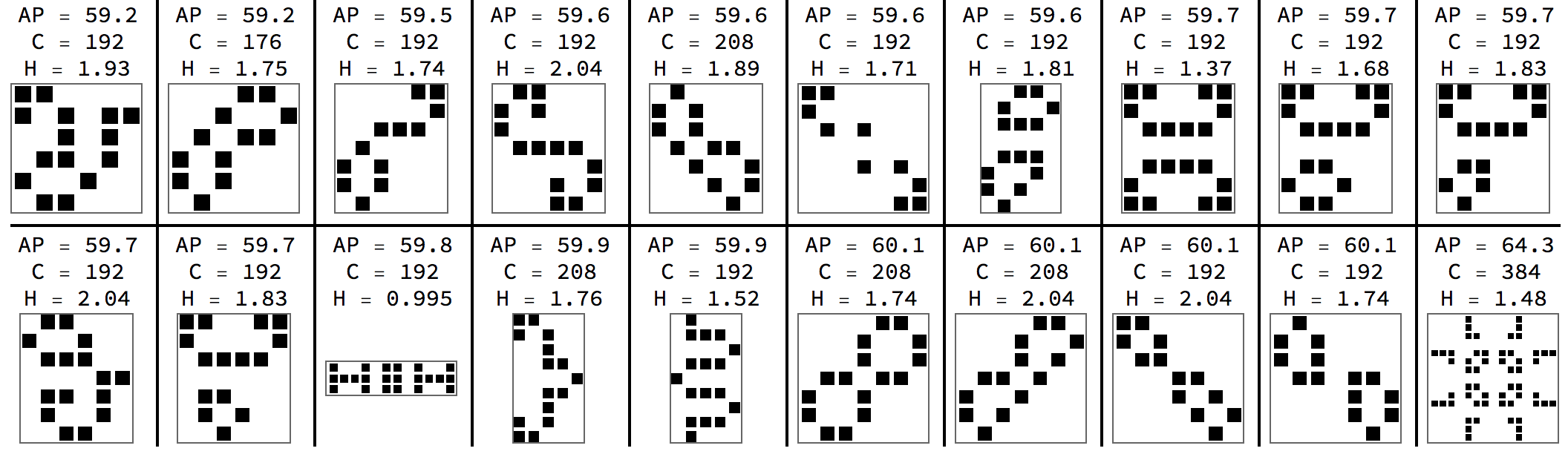}
  \caption{\label{topbottom}A: Top 20 and B: bottom 20 most and least algorithmically complex local persistent patterns in GoL (AP is the BDM estimation, C is lossless compression by Compress, and H is classical Shannon Entropy).} 
\end{figure}

Algorithmic probability may not account for a greater percentage of the deviation from uniform or normal distribution because patterns are filtered by persistence, i.e. only persistent patterns are retained after an arbitrary runtime step, and therefore no natural halting state exists, likely producing a difference in distribution as reported in~\cite{emergence}, where distributions from halting and non-halting models of computation were studied and analysed. Values of algorithmic probability for some motifs (the top and bottom 20 motifs in GoL) are given in Figure~\ref{topbottom}.

\begin{figure}[ht]
  \centering
  A\hspace{5.4cm}B\\
  
  \medskip
  
    \includegraphics[width=5.3cm]{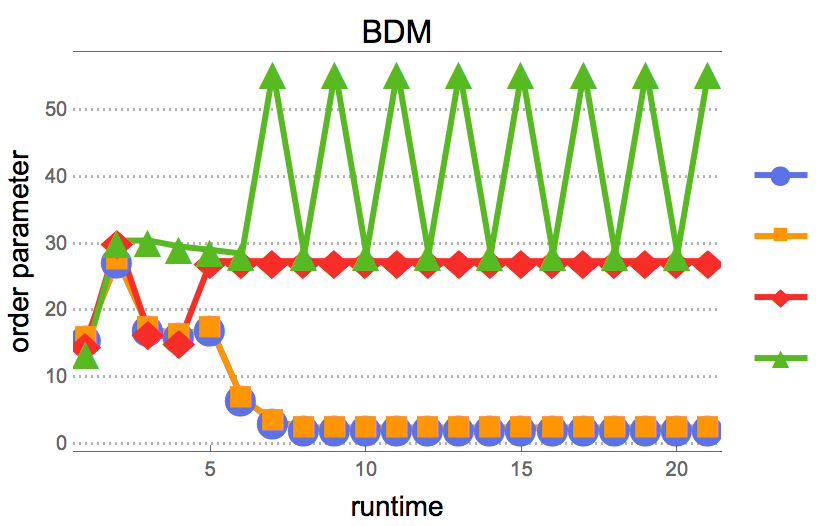}
  \includegraphics[width=6.4cm]{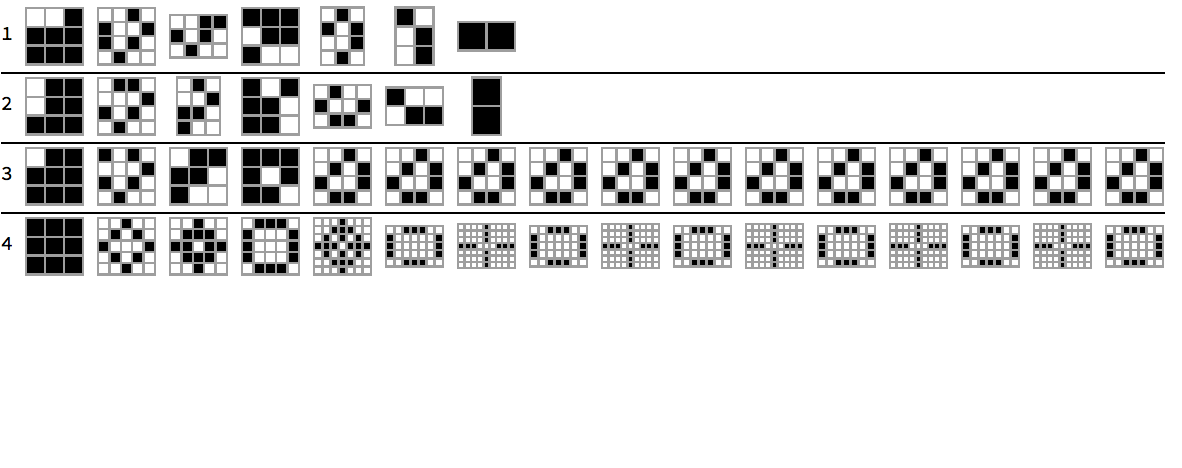}\\
  
    \medskip
    
    C\hspace{5cm}D\hspace{2cm}\\
  
   \includegraphics[width=5.4cm]{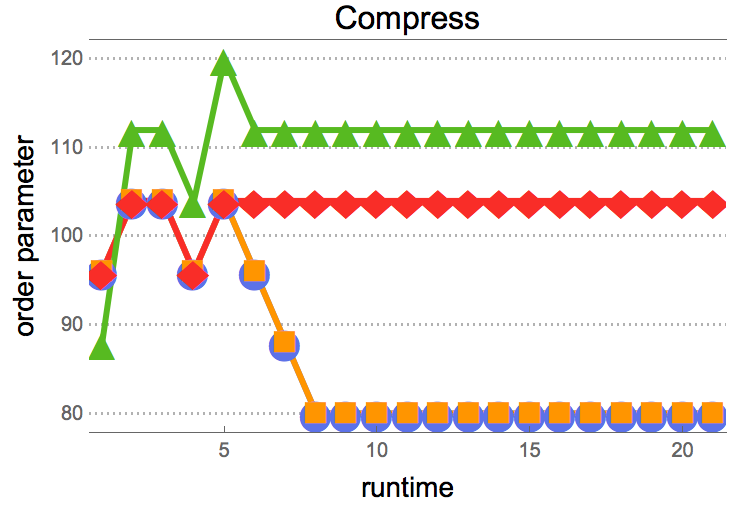}
    \includegraphics[width=5.3cm]{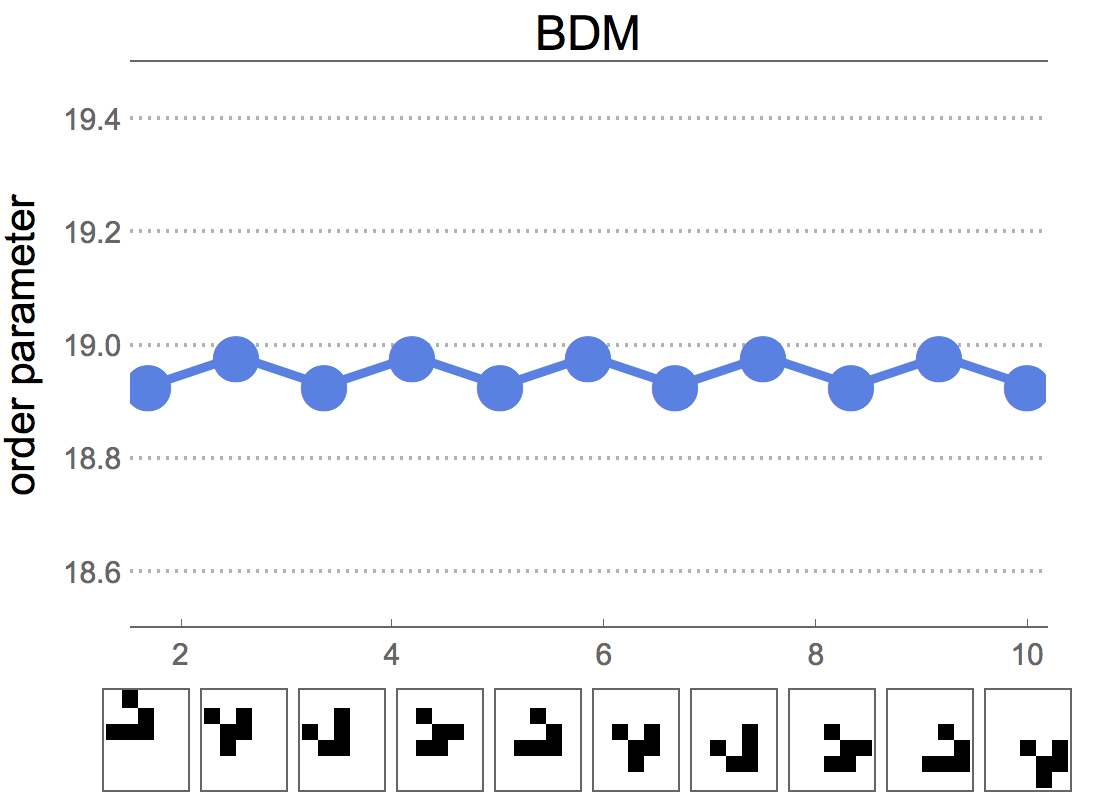}\hspace{1cm}
\caption{\label{GoLcomps3}A: Algorithmic probability approximation of local GoL orbits by BDM on evolving patterns of size $3 \times 3$ cells/pixels in GoL that remain `alive'. B: Same behavioural analysis using Compress (based on LZW) under-performing (compared to BDM) in the characterisation of small changes in local emergent patterns. D: The algorithmic dynamics of a free particle (the most popular local moving pattern in GoL, the glider), with BDM capturing its 2 oscillating shapes in a closed moving window of $4 \times 4$ cells running for 11 steps.} 
\end{figure}

On the other hand, as plotted in Figure~\ref{dist}B-D, the frequency and algorithmic complexity of the patterns in GoL follow a rank distribution and are negatively correlated amongst each other, just as the algorithmic Coding theorem establishes. That is, the most frequent emergent patterns are also the most simple, while the most seldom are more algorithmically random (and their algorithmic probability low). This is also illustrated by plotting the complexity of the distribution of patterns in GoL as they emerge with long tails for both still and periodic patterns and for all patterns of increasing square window size.

\begin{figure}[htbp!]
  \centering
A\hspace{5.7cm}B\\

\medskip

\includegraphics[width=5cm]{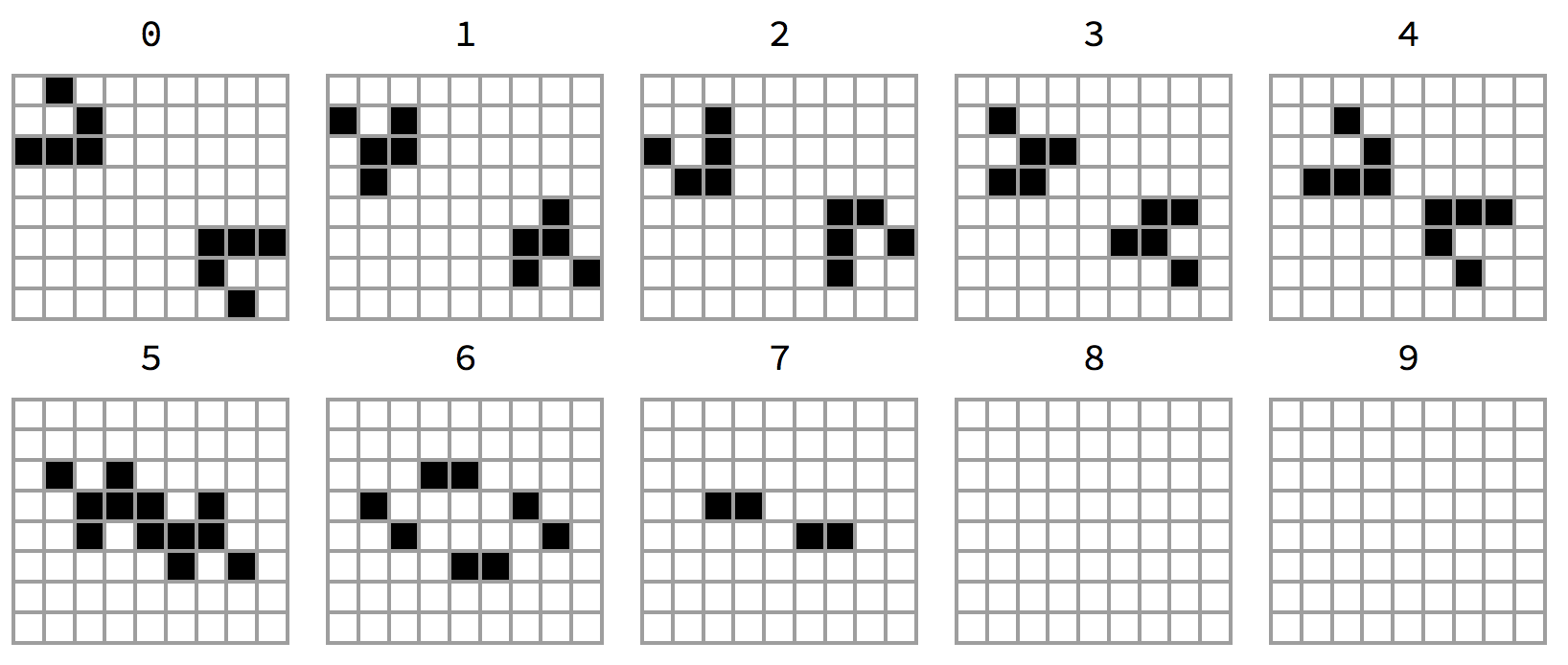}\hspace{2cm}
\includegraphics[width=3.7cm]{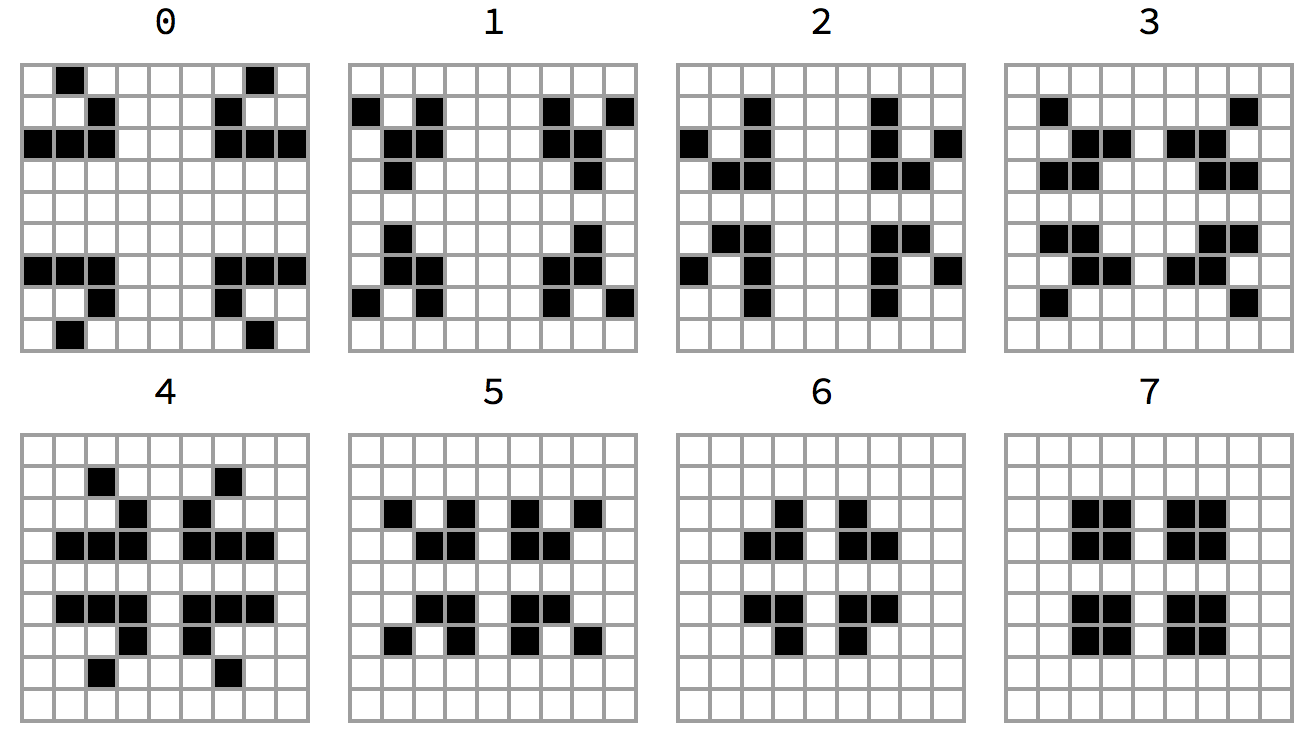}\\

C\hspace{6cm}\\

\medskip

\includegraphics[width=7cm]{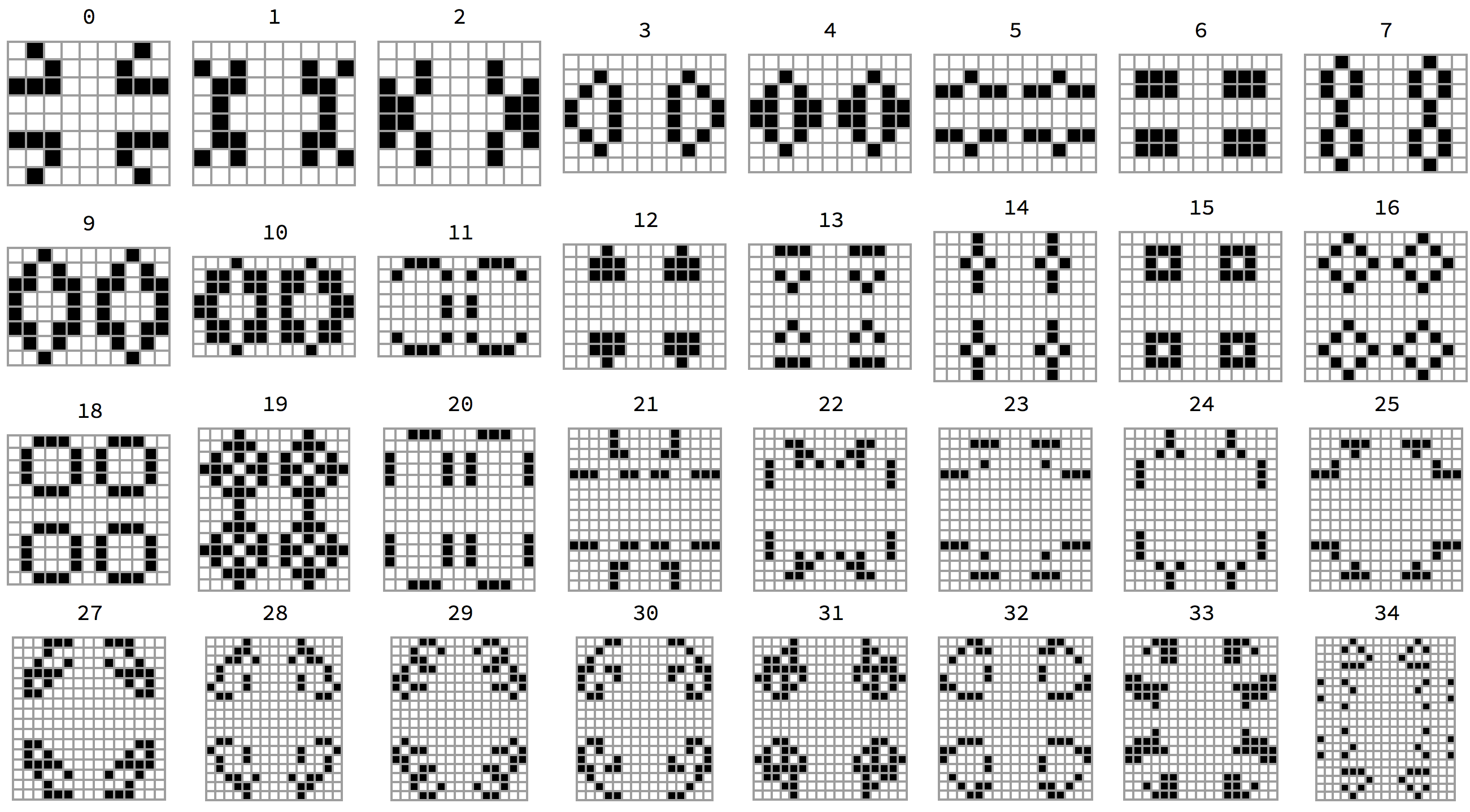}
\\

\medskip

D\hspace{6cm}\\

\includegraphics[width=7cm]{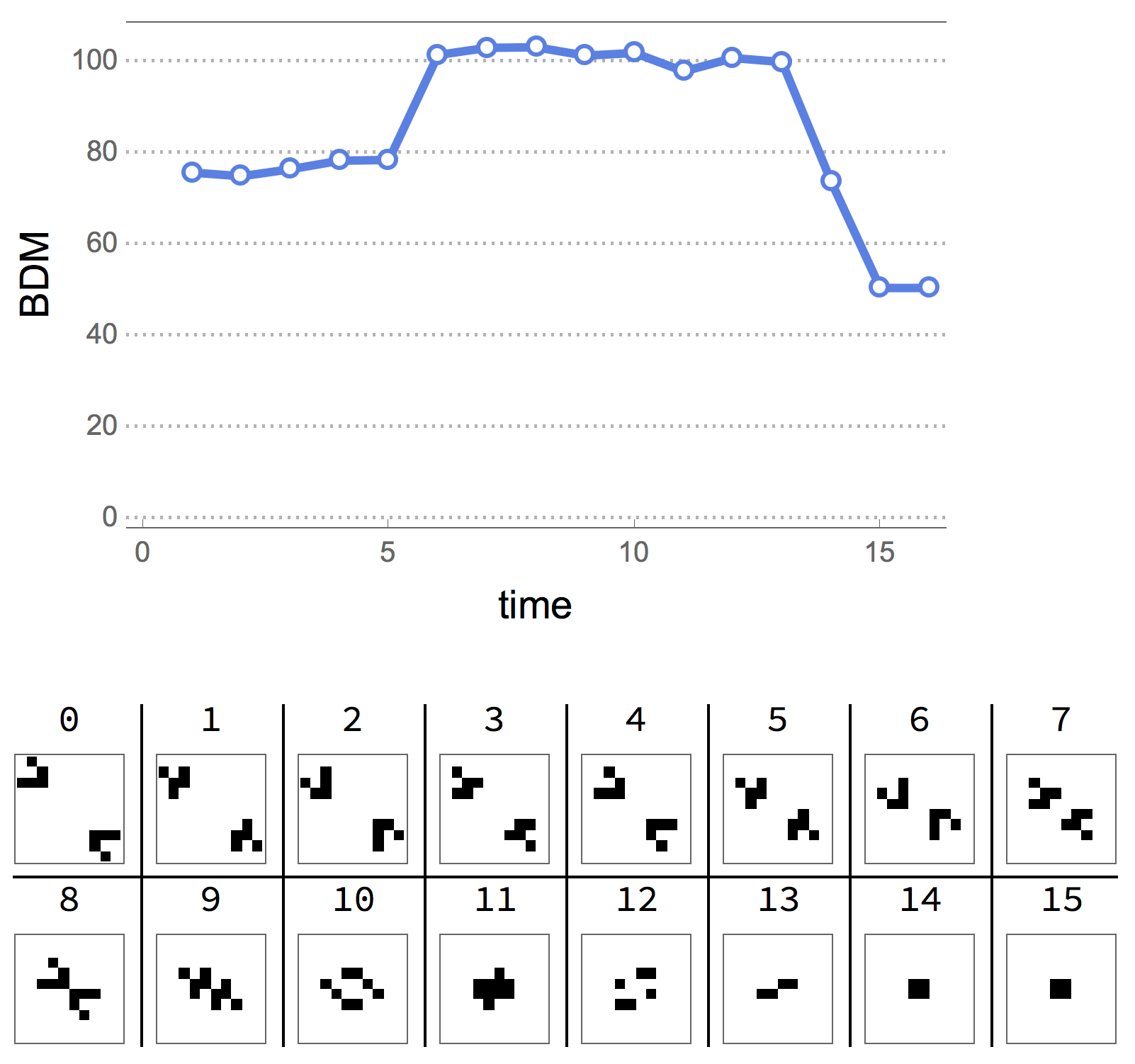}\\
  
  \caption{\label{collision1}A, B and C: 3 possible collisions showing 2-particle annihilation (A), stability (B) and instability, i.e. production of new particles (C). D: The algorithmic information dynamics of a 2-particle stable collision.} 
\end{figure}

\begin{figure}[ht]
  \centering
    A\hspace{6.5cm}B\hspace{2cm}\\
  
    \medskip
    
  \includegraphics[width=5.3cm]{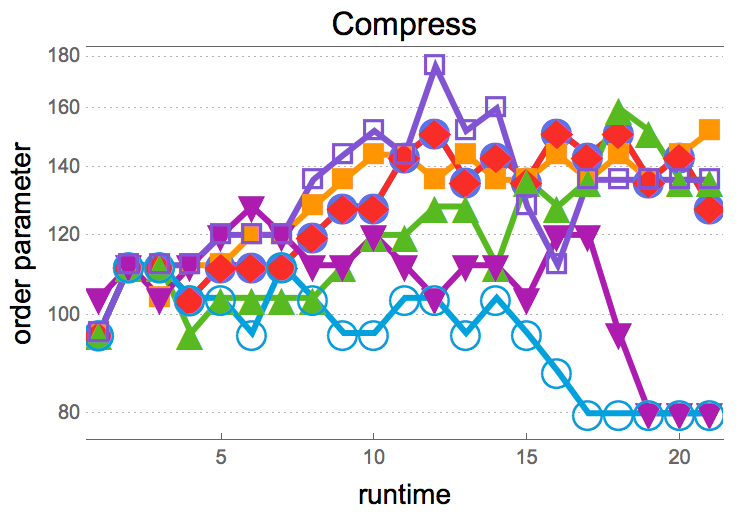}\hspace{1.5cm}\includegraphics[width=5.3cm]{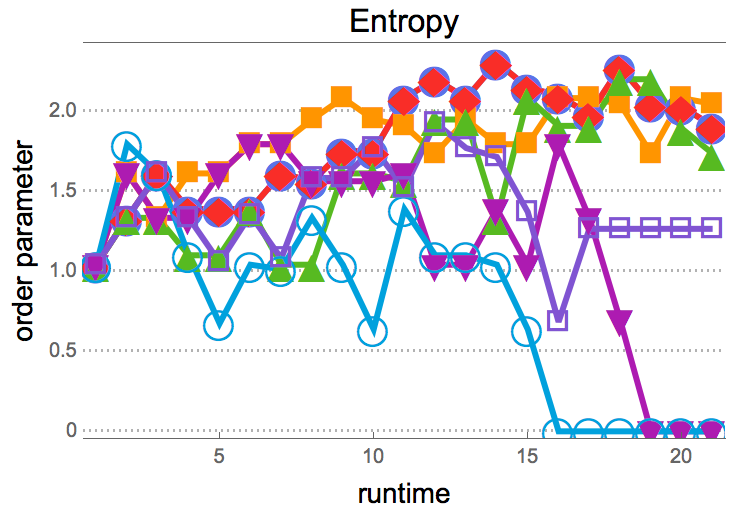}\\
    C\hspace{6.5cm}D\hspace{4cm}\\
  
  \medskip
 
  \includegraphics[width=5.3cm]{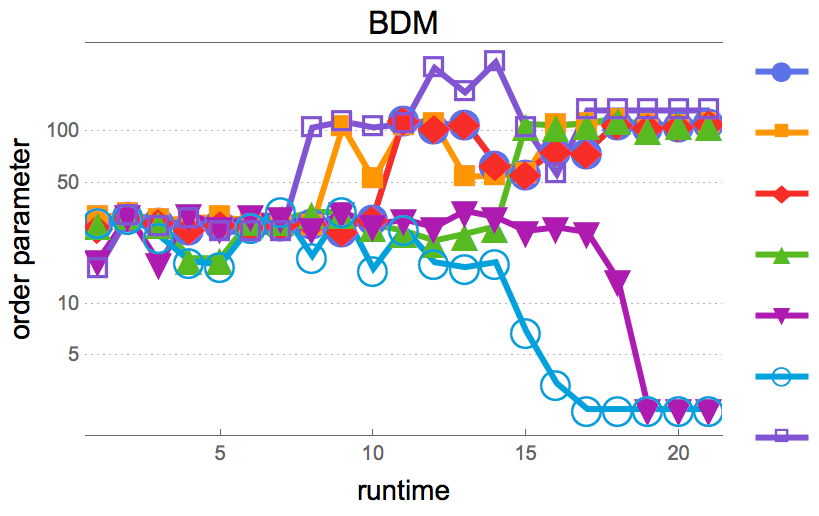}
  \includegraphics[width=6.4cm]{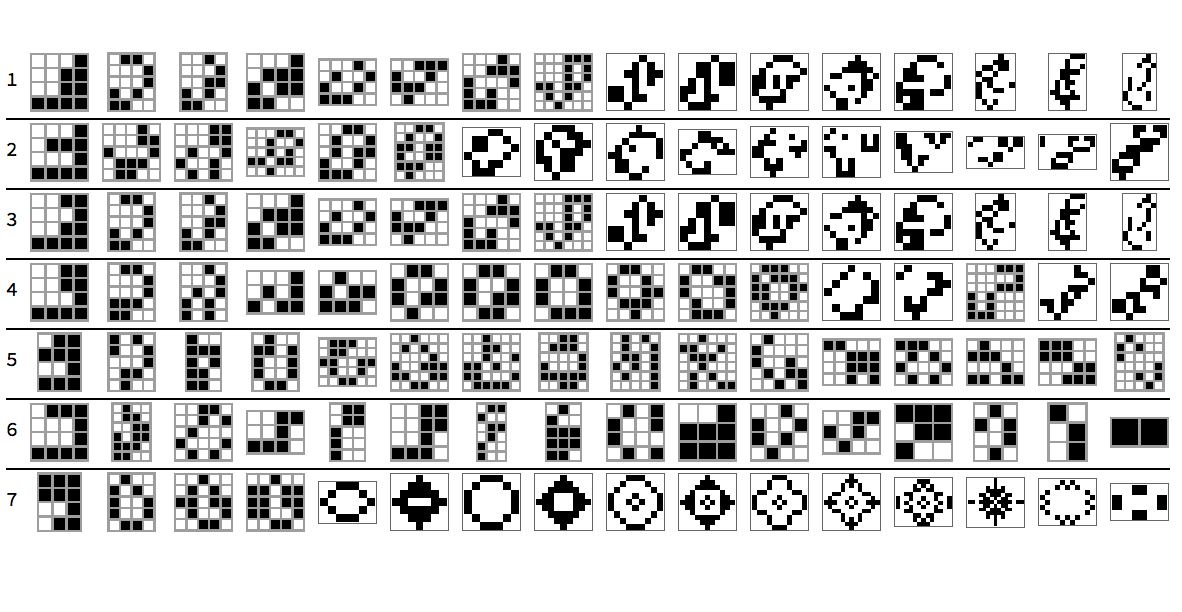}
  \caption{\label{GoLcomps4}Orbit algorithmic dynamics of local emergent patterns in GoL. Compress (A) and Entropy (B) retrieve very noisy results compared to BDM (C) which converges faster and separates the dynamic behaviour of all emerging patterns in GoL of size $4 \times 4$ pixels.}
\end{figure}

\subsection{Algorithmic dynamics of evolving patterns}

While each pattern in GoL evolving in time $t$ comes from the same generating global rule for which $K(GoL(t))$ is fixed (up to $\log(t)$ corresponding to the binary encoding of the runtime step), a pattern within an observational window (Fig.~\ref{obs}) that does not necessarily display the action of all the local rules of the global rule can be regarded as an (open) system separate from the larger system governed by the global rule. This is similar to what happens in the practice of understanding real-world complex systems to which we only have partial access and where a possible underlying global rule exists but is unknown. 

\begin{figure}[htbp!]
  \centering
A\hspace{4cm}B\\

\medskip

\includegraphics[width=5.8cm]{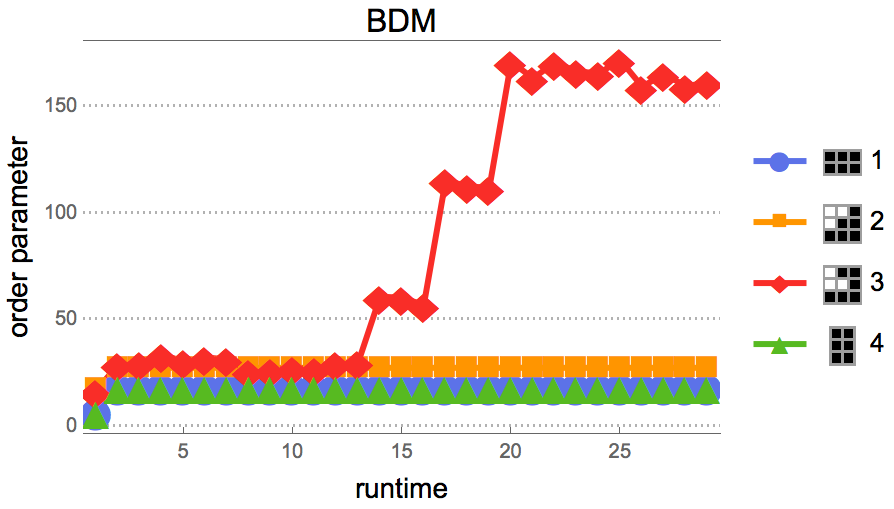}\hspace{1cm}\includegraphics[width=4.8cm]{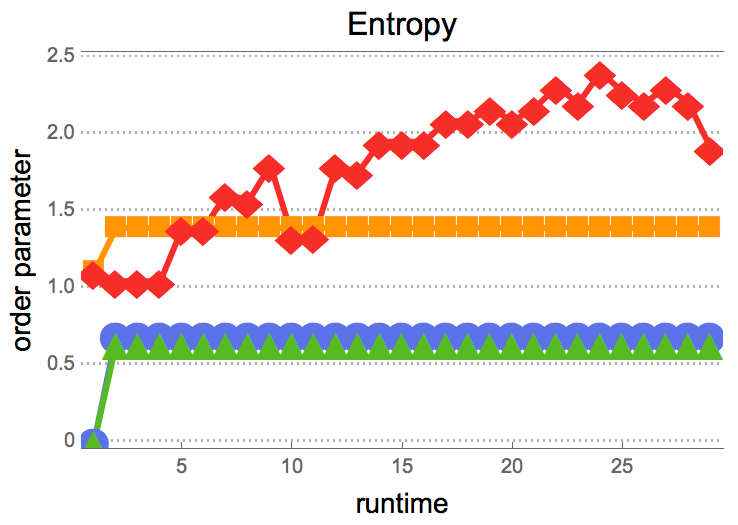}\\
  
C\\

\medskip
 \includegraphics[width=8cm]{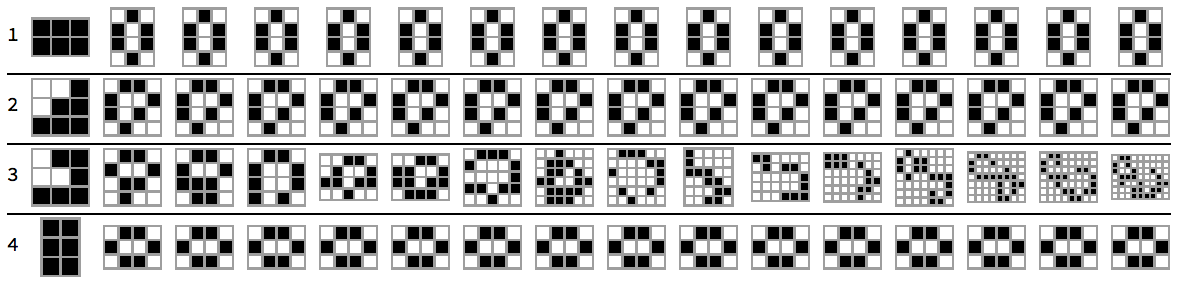}\\
  
  \medskip
  
  D\\

\medskip

  \medskip
  \includegraphics[width=9.5cm]{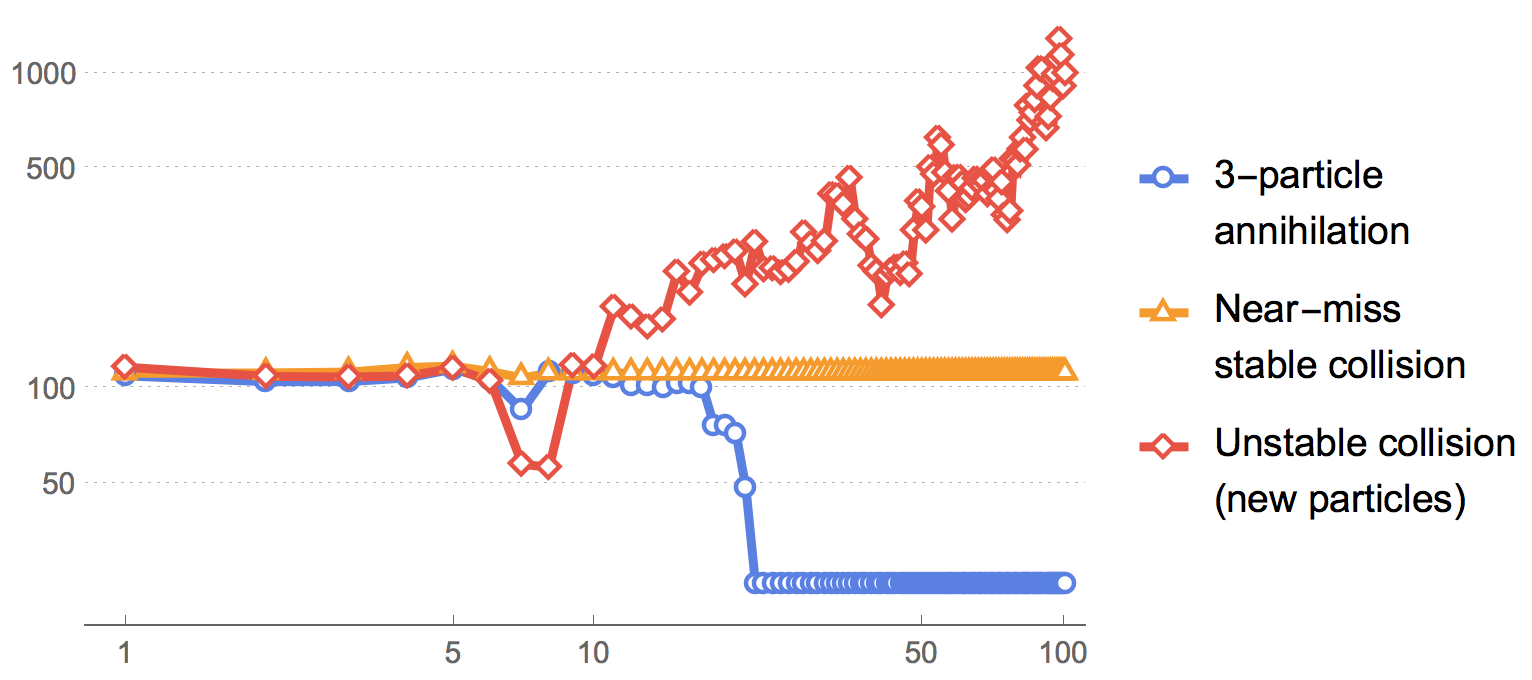}

  \caption{Orbit complexity profiling. A and B: collapsing all the simplest cases (1, 2 and 4) to the bottom, closest to zero, values
  diverging from the only open-ended case (3). A: The measure BDM returns the best separation compared to Entropy C: 16 steps corresponding to evolving steps of the 4 cases captured in A and B. C: The algorithmic information dynamics of 3 particle interactions/collisions. The unstable collision corresponds to Figure~\ref{collision1}D, the 3-particle annihilation is qualitatively similar to the 2-particle Figure~\ref{collision1}A and the near-miss stable collision corresponds to Figure~\ref{collision1}B where the 4 particles look as if about to collide but appear not to (hence a `near miss'). Starting seeds are shown in (see Figure~\ref{seeds} Sup. Inf.).} 
  \label{Golruns2}
\end{figure}

An application to Conway's Game of Life evolving over time using BDM shows some advantages in the characterisation of emergent patterns, as seen in Figures~\ref{GoLcomps3},\ref{GoLcomps4}, and~\ref{Golruns2}.

\begin{figure}[ht]
  \centering
A\hspace{6cm}\\

\medskip

\includegraphics[width=11.5cm]{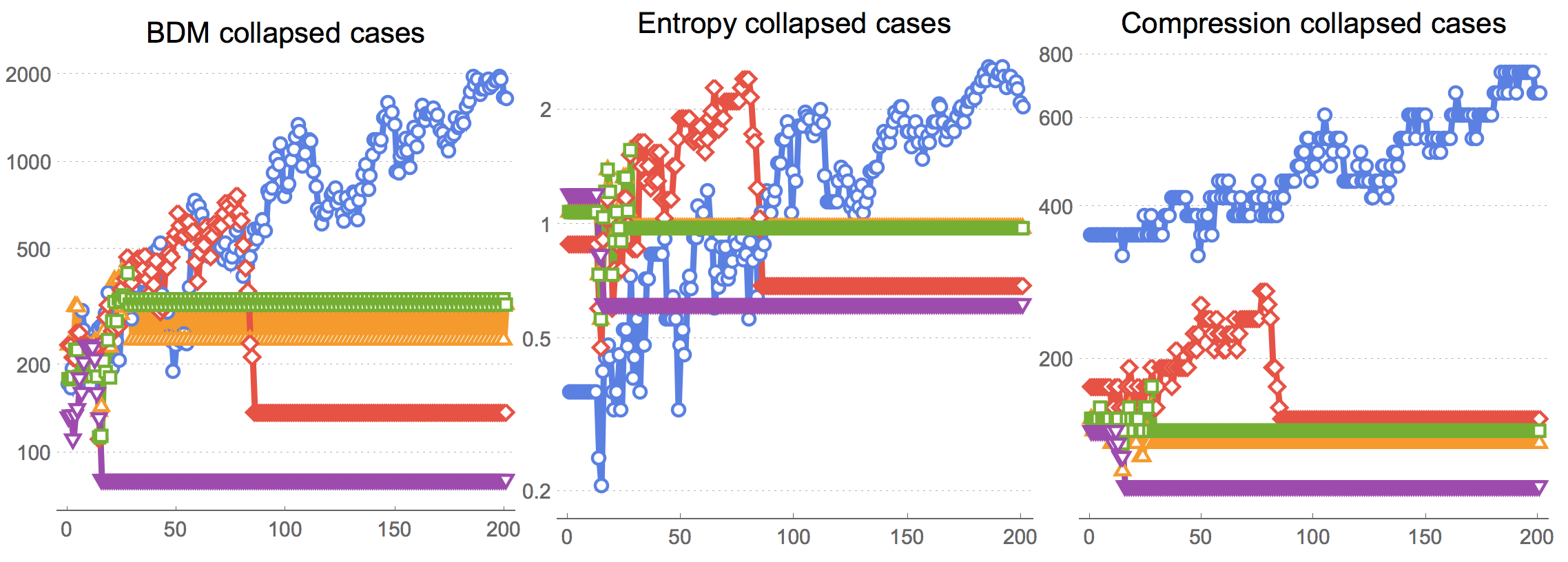}\\

\medskip

B\hspace{6cm}\\

\includegraphics[width=11.2cm]{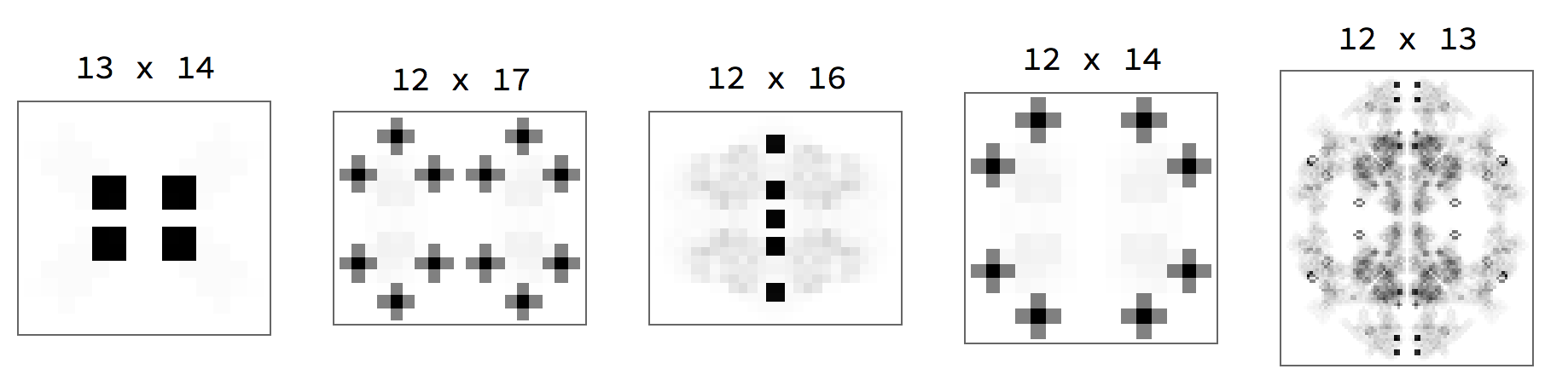}\\
  
\caption{\label{collapsedcases}A: Collapsed cases suggesting clusters of dynamical system attractors of colliding gliders in GoL. B: Density plot of all non-trivial (particles that are not entirely annihilated)
qualitative interactions among 4 particles. The darker the later and more persistent in time.} 
  
\end{figure}

We took a sliding window consisting of a small number of $n \times m$ cells from a 2D cross section of the 3D evolution of GoL as shown in Figure~\ref{obs}. For most cases $n=m$. The size of $n$ and $m$ is determined by the size of the pattern of interest, with the sliding window following the unfolding pattern. The values of $n$ or $m$ may increase if the pattern grows but never decreases, even if the pattern disappears. Each line in all plots corresponds to the algorithmic dynamics (complexity change) of the orbit of a local pattern in GoL, unless otherwise established (e.g. such as in collapsed cases). Figure~\ref{GoLcomps3}, for example, demonstrates how the algorithmic probability approach implemented by BDM can capture dynamical changes even for small patterns where lossless compression may fail because limited to statistical regularities that are not always present. For example, in Figure~\ref{GoLcomps3}A, BDM captures the periodic/oscillating behaviour (period 2) of a small pattern, something that compression, as an approximation to algorithmic complexity, was unable to capture for the same motifs in Figure~\ref{GoLcomps3}B. Likewise, the BDM approximation to algorithmic complexity captures the periodic behaviour of the glider in Figure~\ref{GoLcomps3}D for 10 steps.

Figures~\ref{GoLcomps4}A and B illustrate cases of diagonal particle (glider) collisions. In a slightly different position, the same 2 particles can produce a single still pattern as shown in Figure~\ref{GoLcomps4}D, that reaches a maximum of complexity when new particles are produced, thereby profiling the collision as a transition between a dynamic and a still configuration. In Figure~\ref{GoLcomps4}A the particles annihilate each other after a short transition of different configurations. In Figure~\ref{GoLcomps4}B the collision of 4 gliders produces a stable non-empty configuration of still particles after a short transition of slightly more complicated interactions. We call this interaction a `near-miss' because the particles seem to have missed each other even though there is an underlying interaction. In Figure~\ref{GoLcomps4}C, an unstable collision characterised by the open-ended number of new patterns evolving over time in a growing window can also be characterised by their algorithmic dynamics using BDM, as shown in Figure~\ref{Golruns2}D and marked as an unstable collision.

More cases, both trivial and non-trivial, are shown in Figures~\ref{GoLcomps4} and ~\ref{Golruns2}A and B. Figure~\ref{GoLcomps4} shows other 7 cases of evolving motifs starting from different initial conditions in small grid sliding windows of size up to 4$\times$4 displaying different evolutions captured by their algorithmic dynamics. Figure~\ref{Golruns2} shows all evolving patterns of size 3$\times$3 in GoL and the algorithmic dynamics characterising each particle's behaviour, with BDM and Entropy showing similar results, but a better separation for BDM.

\subsection{Algorithmic dynamic profiling of particle collisions}

We traced the evolution of collisions of so-called gliders. Figure~\ref{collision1} Sup. Inf. shows concrete examples of particle collisions of gliders in GoL and the algorithmic dynamic characterisation of one such interaction, and Figure~ \ref{collapsedcases}A illustrates all cases for a sliding window of up to size $17 \times 17$ where all cases for up to 4 colliding gliders are reported, analysed and classified by different information-theoretic indexes, including compression as a typical estimator of algorithmic complexity and BDM as an improvement on both Shannon entropy alone and typical lossless compression algorithms. The results show that cases can be classified in a few categories
corresponding to the qualitative behaviour of the possible outcomes of up to 4 particle collisions.

Figure~\ref{Golruns2}D summarises the algorithmic dynamics of different collisions and for all cases with up to 4 gliders in Figure~\ref{collapsedcases}A by numerically producing all collisions but collapsing cases into similar behaviour corresponding to qualitatively different cases, as shown in the density plots in  Figure~\ref{collapsedcases}B. The interaction of colliding particles is characterised by their algorithmic dynamics, with the algorithmic probability estimated by BDM remaining constant in the case in which 4 particles prevail, the annihilation case collapsing to 0, and the unstable collision producing more particles diverges.

\section{Conclusions}

We have explained how observational windows can be regarded as apparently open systems even if they come from a closed deterministic system $D(t)$ for which the algorithmic complexity $K(D)$ cannot differ by more than $\log_2(t)$ over time $t$--a (mostly) fixed algorithmic complexity value. 
However, in local observations patterns seem to emerge and interact, generating algorithmic information as they unfold and requiring different local rules and revealing the underlying mechanisms of the larger closed system.

We have shown the different capabilities that both classical information and algorithmic complexity (the former 
represented by the lossless compression algorithm Compresss, 
and the latter based on algorithmic probability) display in the characterisation of these objects and how they can be used and exploited to track changes and analyse their spatial dynamics.

We have illustrated the way in which the method and tools of \textit{algorithmic dynamics} can be used and exploited to measure the \textit{algorithmic information dynamics} of discrete dynamical systems, in particular of emerging local patterns (particles) and interacting objects (such as colliding particles), as exemplified in a much studied 2-dimensional cellular automaton. 

\section*{Acknowledgements}

H.Z. was supported by Swedish Research Council (Vetenskapsr\r{a}det) grant No. 2015-05299.

\newpage

\section*{Supplementary Information}

\begin{figure}[htbp!]
  \centering
  \includegraphics[width=1.5cm]{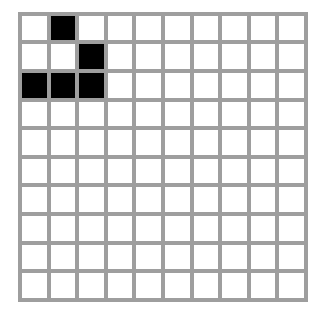}\hspace{.5cm} \includegraphics[width=1.5cm]{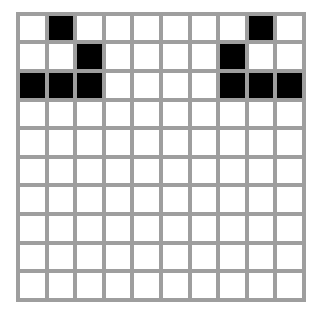}\hspace{.5cm}
   \includegraphics[width=1.5cm]{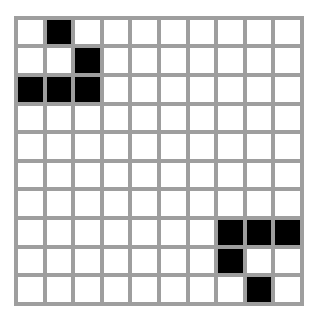}\hspace{.5cm}
    \includegraphics[width=1.5cm]{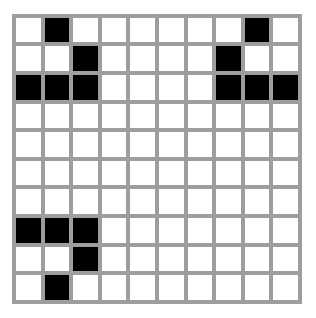}\hspace{.5cm}
     \includegraphics[width=1.5cm]{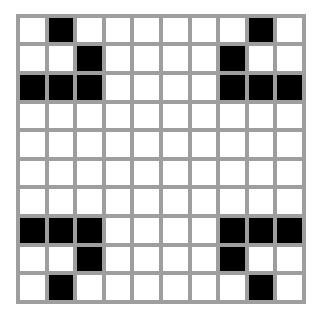}
  \caption{\label{seeds}Set of initial conditions set for particle (glider) collision. From left to right: free particle, 2-particle sideways     collision, 2-particle frontal collision, 3-particle collision and 4-particle collision.} 
\end{figure}

\end{document}